\documentclass{aa}
\usepackage{graphicx}
\def\etal{{\it et~al.~}}
\def\nt{nonthermal~}
\def\bsax{{\it BeppoSAX~}}
\def\ginga{{\it Ginga~}}
\def\einstein{{\it Einstein~}}

\def\asca{{\it ASCA~}}
\def\rosat{{\it ROSAT~}}
\def\rxte{{\it RXTE~}}

\def\erg{~{\rm erg~ cm}^{-2}\ {\rm s}^{-1}~}
\def\ergs{~{\rm erg~s}^{-1}~}
\def\h0{~{\rm H_0 = 50~km~s}^{-1}\ {\rm Mpc}^{-1}~h_{50}~}
\def\gtsim{\raise 2pt \hbox {$>$} \kern-1.1em \lower 4pt \hbox {$\sim$}}

\begin{document}
\title{Hard X-ray and radio observations of \\ Abell 754}
\author{R.Fusco-Femiano\inst{1} \and M.Orlandini\inst{2} \and S.De Grandi\inst{3} \and
S.Molendi\inst{4} \and L.Feretti\inst{5} \and
G.Giovannini\inst{5,6}, \\M.Bacchi\inst{5,6} \and
F.Govoni\inst{5,6}}

\offprints{R. Fusco-Femiano}

\institute{Istituto di Astrofisica Spaziale e Fisica Cosmica
(IASF), CNR, via del Fosso del Cavaliere 100, 00133
 Roma, Italy (dario@rm.iasf.cnr.it) \and IASF, CNR, Bologna,
Italy \and Osservatorio Astronomico di Brera, Merate, Italy \and
IASF, CNR, Milano, Italy \and Istituto di Radioastronomia, CNR,
Bologna, Italy \and Dip. di Astronomia, Univ. di Bologna, Bologna,
Italy}
\date{Received 2 July 2002 / Accepted 3 November 2002}
\abstract{We present a long \bsax observation of Abell 754 that
reports a nonthermal excess with respect to the thermal emission
at energies greater than $\sim$ 45 keV. A VLA radio observation at
1.4 GHz definitely confirms the existence of diffuse radio
emission in the central region of the cluster, previously
suggested by images at 74 and 330 MHz (Kassim \etal 2001), and
reports additional features. Besides, our observation determines a
steeper radio halo spectrum in the 330-1400 MHz frequency range
with respect to the spectrum detected at lower frequencies,
indicating the presence of a spectral cutoff. The presence of a
radio halo in A754, considered the prototype of a merging cluster,
reinforces the link between formation of Mpc-scale radio regions
and very recent or current merger processes. The radio results
combined with the hard X-ray excess detected by \bsax give
information on the origin of the electron population responsible
for nonthermal phenomena in galaxy clusters. We discuss also the
possibility that 26W20, a tailed radio galaxy with BL Lac
characteristics located in the field of view of the PDS, could be
responsible for the observed nonthermal hard X-ray emission.

\keywords{Extragalactic Astronomy -- Clusters of Galaxies -- Radio
and X-ray Astronomy }}

\maketitle

\section{Introduction}
Shocks and turbulence associated with a major cluster merger event
could provide the ingredients necessary to the formation of
extended radio regions detected so far in a limited number of
clusters, namely a diffuse magnetic field amplification and
particle reacceleration (Tribble 1993; Roettiger, Burns, \& Stone
1999; Roettiger, Stone, \& Burns 1999; Brunetti \etal 2001). The
existence of Mpc-scale radio halos or relics combined with the
relatively short radiative lifetimes suggests an in-situ electron
reacceleration induced by a very recent or current merger event.
This hypothesis is supported by observational evidence (Feretti
1999). The existence of large radio regions could be at the origin
of the nonthermal hard X-ray (HXR) emission detected in the Coma
cluster (Fusco-Femiano \etal 1999; Rephaeli, Gruber, \& Blanco
1999) and Abell 2256 (Fusco-Femiano \etal 2000) both showing
extended radio emission. A considerable fraction of the input
energy during a merger process can be released in particle
acceleration and re-emitted in various energy bands. In
particular, the same accelerated electrons responsible for the
diffuse radio synchrotron emission can scatter the Cosmic
Microwave Background (CMB) photons to produce nonthermal inverse
Compton (IC) X-ray emission. Alternative relativistic particles to
primary re-accelerated electrons emitting in radio halos or relics
and responsible for IC nonthermal HXR emission could be secondary
electrons produced in the interactions of cosmic rays in the ICM
(Dennison 1980; Blasi \& Colafrancesco 1999).

X-ray and optical observations report a violent merger event in
the galaxy cluster Abell 754 (Henry \& Briel 1995; Zabludoff \&
Zaritsky 1995; Henriksen \& Markevitch 1996; Bliton \etal 1998;
Markevitch \etal 1998; De Grandi \& Molendi 2001) and a numerical
hydro/N-body model of this cluster (Roettiger, Stone, \& Mushotzky
1998) has shown that many of its observed morphological properties
can be explained by a very recent merger ($<$ 0.3 Gyr), slightly
off-axis, between two clusters having a total mass-ratio less than
2.5:1. Therefore, the intracluster medium (ICM) of A754 appears to
be a suitable place for the formation of radio halos or relics. As
a consequence, radio and HXR observations of this cluster are
relevant to establish robust evidence of the suggested link
between the presence of nonthermal phenomena and merger activity
in clusters of galaxies and to obtain information on the processes
that can inject relativistic electrons. The cluster has been
recently imaged at 74 and 330 MHz with the NRAO VLA observatory
(Kassim \etal 2001) suggesting the existence of a radio halo and
at least one radio relic. Observations at higher resolution and
sensitivity are required to confirm the obtained results. A754 was
observed in hard X-rays with \rxte in order to search for a
nonthermal component (Valinia \etal 1999). The fit to the PCA and
HEXTE data set an upper limit of $\sim$ 1.4$\times$ 10$^{-12}\erg$
in the 10-40 keV band to the nonthermal emission.

In this paper we present the results of a high sensitivity VLA
radio observation at 1.4 GHz and of a long \bsax observation,
exploiting the unique capabilities of the PDS (Frontera \etal
1997) to search for HXR emission. We discuss the new radio data
and the origin of the excess detected at energies above $\sim$ 45
keV with respect to the thermal emission.

Throughout this paper, we assume a Hubble constant of $\h0$ and
$q_0$ = 1/2, so that an angular distance of $1^{\prime}$
corresponds to 86 kpc ($z_{A754}$ = 0.054; Bird 1994). Quoted
confidence intervals are at a 90\% level, if not otherwise
specified.

\section{Hard X-ray data}

\subsection{PDS and MECS data reduction}

The pointing coordinates of \bsax are at J2000 $\alpha$ = 9$^h$
9$^m$ 21.0$^s$, $\delta$ = -9$^{\circ}$ 41$^{\prime}$ 24.0$^{"}$.
The total effective exposure time for the PDS was $\sim$ 84250 sec
in the two observations of 2000 May 6$^{th}$ and May 17$^{th}$.
The observed count rate was 0.333 $\pm$ 0.021 counts s$^{-1}$ in
the 15-100 keV energy range, at a confidence level of
$\sim$16$\sigma$.

Since the source is rather faint in the PDS band (approximately 2
mcrab in the 15-100 keV), a careful check of the background
subtraction must be performed. PDS spectra were extracted using
the XAS v2.0 (Chiappetti \& Dal Fiume 1997). The background
sampling was performed using the default rocking law of the two
PDS collimators that samples ON,+OFF, ON,-OFF fields for each
collimator with a dwell time of 96 s (Frontera \etal 1997). When
one collimator is pointing ON source, the other collimator is
pointing toward one of the two OFF positions. We used the standard
procedure to obtain PDS spectra (Dal Fiume \etal 1997); this
procedure consists of extracting one accumulated background
spectra in the two different +/-OFF sky direction. The comparison
between the two accumulated backgrounds ([+OFF]vs.[-OFF]) shows a
difference in the pointings for the first of the two observations.
In particular, the background in the position +OFF was much
greater than that in position -OFF. So, for the analysis, we have
considered only the background in the -OFF position. In the second
observation we do not find differences between the spectra
extracted from the two offset positions. The background level of
the PDS is the lowest and most stable obtained so far with the
high-energy instruments on board satellites thanks to its
equatorial orbit. No modelling of the time variation of the
background is required. The correctness of the PDS background
subtraction has been checked by verifying that the counts
fluctuate at about zero flux as the signal falls below
detectability. This happens at energies greater than $\sim$ 75
keV.

The total exposure time for the MECS was $\sim$ 1.85$\times$10$^5$
sec with a count rate of 0.748 $\pm$ 0.002 counts s$^{-1}$. MECS
data preparation and linearization were performed using the SAXDAS
package under an FTOOLS environment. We have extracted an MECS
spectrum from a circular region of $\sim$ 20$^{\prime}$ radius
(corresponding to about 1.7 Mpc). From the \rosat PSPC radial
profile, we estimate that about 95\% of the total cluster emission
falls within this radius. The background subtraction has been
performed using spectra extracted from blank sky event files in
the same region of the detector as the source. For more details
see De Grandi \& Molendi (2001) that used the same MECS data to
determine the temperature profile of A754.

A numerical relative normalization factor among the two
instruments has been included in the fitting procedure (see next
section) to account for (1) the fact that the MECS spectrum
included emission out to $\sim$ 1.7 Mpc from the X-ray peak, while
the PDS field of view (FWHM = 1.3$^{\circ}$) covers the entire
emission from the cluster; (2) the slight mismatch in the absolute
calibration of the MECS and PDS response matrices employed; and
(3) the vignetting in the PDS instrument (the MECS vignetting is
included in the response matrix). The estimated normalization
factor is $\sim$ 0.8. In the fitting procedure, we allow this
factor to vary within 15\% from the above value to account for the
uncertainty in this parameter.

\begin{figure*}
\centering
\resizebox{10cm}{!}{\includegraphics[angle=-90]{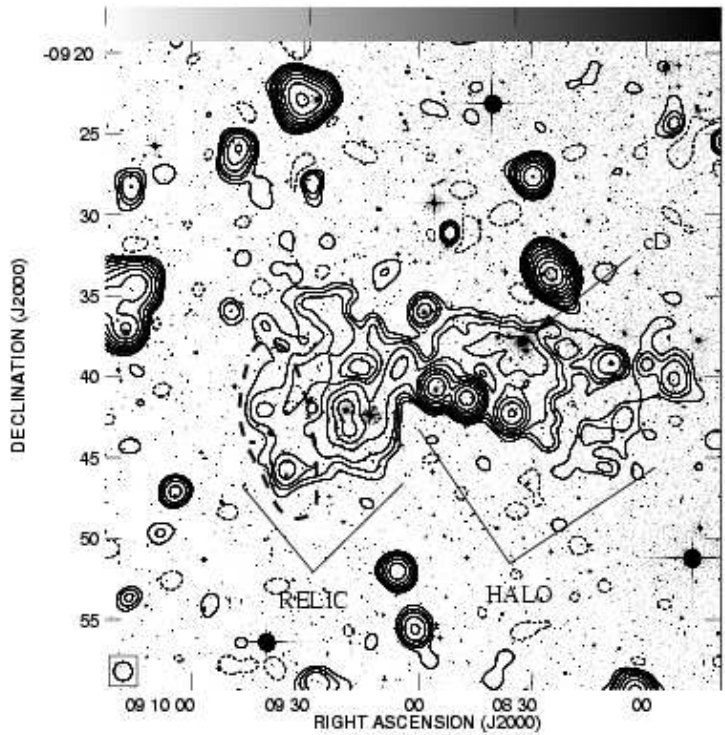}}
\centering
\parbox[b]{80mm}{\caption{Radio map at 20 cm
with resolution of 70\arcsec (contours), overlayed onto the
grey-scale image  from the digitized ESO/SERC Southern Sky Survey.
The $\sigma$ noise level in this map is 0.1 mJy/beam. Contours are
at --0.3, 0.3, 0.6, 1, 2, 4, 8, 16, 32, 64, 128 mJy/beam The
cluster cD galaxy is marked by an arrow. The cluster halo and
relics are indicated. The bold dashed contour indicates the east
relic of Kassim \etal (2001) at 74 MHz.} \label{pip}}
\end{figure*}

\section{Radio data}

A radio observation at 1.365 GHz was obtained with the Very Large
Array (VLA) in D configuration on September 25, 2000, for a total
integration time of about 2.5 h. Data were calibrated and reduced
with the Astronomical Image Processing System (AIPS), following
the standard procedure (Fourier inversion, Clean and Restore,
Self-calibration).

The radio image obtained with angular resolution of 70\arcsec~ is
presented in Fig.1, overimposed onto the optical image.  The rms
noise is of 0.1 mJy/beam, i.e. about 5 times better than the NRAO
VLA Sky Survey (NVSS) image (published by Giovannini \etal 1999).
The diffuse emission detected here is complex and it is much more
extended than in any previous image at any frequency.

The cluster halo is easily visible around the brightest optical cD
galaxy. This emission is not detected in the NVSS, but is detected
at 330 MHz and 74 MHz by Kassim \etal (2001). As pointed out by
these authors, several discrete sources are embedded within the
radio halo. In addition, we detect in the present image an
extended diffuse emission to the east of the halo. Although a
connection of this emission to the radio halo cannot be excluded,
we tentatively classify this feature as a peripheral cluster
relic. The cluster radio relic is detected by Kassim \etal (2001)
at 74 MHz and not at 330 MHz. At 74 MHz it is much less extended
than the feature detected here at 1.4 GHz, and it is located at
the eastern boundary of our relic image. The possible connection
relic-halo present in the 1.4 GHz image does not appear in the 74
MHz image.

The total flux density of the radio halo at 1.4 GHz, after
subtraction of the discrete sources, is 86 $\pm$ 2 mJy. Comparison
to the total flux at 330 MHz given by Kassim \etal (2001) leads to
a spectral index $\alpha^{1400}_{330}$ $\sim$ 1.5. The halo
spectrum appears to be steeper than that estimated by the previous
observations in the 74-330 MHz frequency band ($\alpha^{330}_{74}$
$\sim$ 1.1), indicating the presence of a spectral cutoff. The
total flux density of the relic is 69 $\pm$ 2 mJy. Since the relic
source in the map of Kassim \etal (2001) at 74 MHz is
significantly less extended than in the present map at 1.4 GHz, we
obtain a lower limit $\sim$ 1 for the spectral index in the range
74-1400 MHz. An upper limit of $\sim$ 1.7 in the range 330-1400
MHz is derived from the lack of detection in the 330 MHz image of
Kassim \etal (2001).

\subsection{Hard X-ray results}

The temperature profile derived from the MECS data analysis of
this \bsax observation confirms that A754 is an extremely
disturbed and substructured cluster, as reported in previous X-ray
and optical observations. The MECS spectrum is extracted from a
circular region of radius $\sim 20'$ that gives a mean temperature
of 9.42$^{+0.16}_{-0.17}$ keV consistent with the value of $\sim$
9 keV determined from other X-ray instruments (\asca GIS, \ginga
LAC, \rxte PCA) within their respective errors bars. The iron
abundance results to be 0.25$^{+0.04}_{-0.03}$. These results are
reported in De Grandi \& Molendi 2001.

To search for a nonthermal component in the spectrum of A754, we
have performed a simultaneous fit (see Fig. 2) to the MECS and PDS
data with a MEKAL model that reports an average gas temperature of
$\sim$9.4 keV ($\chi^2$=194.4 for 158 d.o.f.). A systematic error
of 1\% was added in quadrature to the statistical uncertainties of
the spectral data of the PDS, on the basis of the results obtained
from routinely performed Crab Nebula observations. The
normalization factor for the two data sets is $\sim$ 0.72. At
energies above $\sim$ 45 keV there is a conspicuous excess with
respect to thermal emission at a confidence level of $\sim
3.2\sigma$. This long \bsax observation is composed of two
observations with a time interval of few days and the separate
data analysis does not evidence significative variability. We
infer a robust indication for the nonthermal nature of the
detected excess by fitting the MECS/PDS data with a 2-T model and
with a T plus a power law model. In the first model one of the
temperatures has the value of $\sim$ 9.4 keV, while the second
temperature assumes unrealistic values greater than 50-60 keV that
strongly support a nonthermal mechanism for the second component.
In the second model the fit, with a $\chi^2$ value of 177.4 for
156 d.o.f., is not able to determine the spectral index, but the
improvement of the $\chi^2$ value with respect to a single thermal
component (MEKAL model) is significant at more than 99.92\%
confidence level, according to the F-test. In the 10 - 40 keV
energy range the fit gives a nonthermal flux of $\sim 1.6\times
10^{-12}\erg$ consistent with the upper limit reported by the
\rxte observation (Valinia \etal 1999).

\begin{figure*}
\centering
  \resizebox{10cm}{!}{\includegraphics[angle=-90]{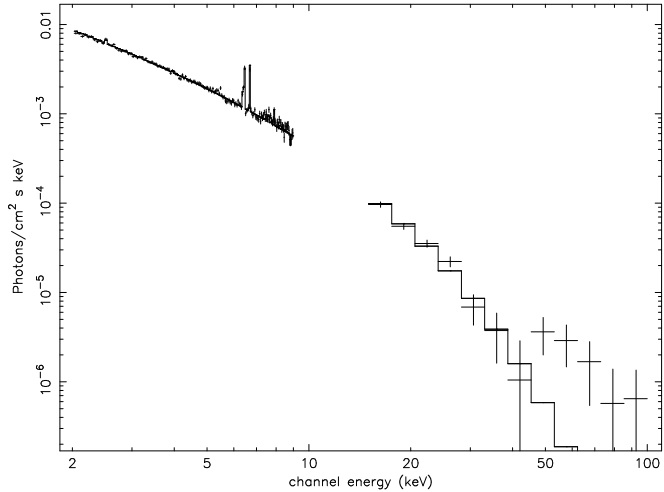}}
  \hfill
  \parbox[b]{80mm}{\caption{MECS data in the 2-9 keV energy range and PDS data
  at energies above 15 keV of the \bsax observation of A754. The continuous line is the fit with a
  MEKAL model. The derived mean gas temperature is of $\sim$9.4 KeV within a region of 20$'$ (1.7 Mpc) of radius
  .}
   \label{pippo}}
\end{figure*}

In the next session we discuss the origin for the detected
nonthermal excess considering the presence of the radio halo in
the ICM of the cluster and that of the radio galaxy 26W20, which
displays BL Lacertae characteristics (Silverman, Harris, \& Junor
1998), in the field of view of the PDS at a distance of about
27$'$ from the \bsax pointing.

\section{Discussion}

Nonthermal X-ray emission has been detected in the Coma cluster
and Abell 2256 by the PDS onboard \bsax, while a flux upper limit
has been reported for Abell 3667 (Fusco-Femiano \etal 2001). The
deviation of these "hard tails" from the the thermal
bremsstrahlung spectra appears to start at energies above 20-25
keV. A likely origin for the detected nonthermal HXR radiation is
IC emission by scattering of CMB photons by the radio synchrotron
electrons responsible for extended radio regions present in the
center (radio halo) or in the periphery (radio relic) of the
cluster. Alternative mechanisms to the IC model have been proposed
(Blasi \& Colafracesco 1999; Sarazin \& Kempner 2000; Dogiel 2000;
Blasi 2000; Liang, Dogiel, \& Birkinshaw 2002), some of these
motivated by the discrepancy between the value for the
intracluster magnetic field derived by the \bsax observation of
the Coma cluster ($B_X\sim 0.16\mu G$; Fusco-Femiano \etal 1999)
and the value derived from the Faraday rotation ($B_{FR}\sim 6\mu
G$; Feretti \etal 1995) of polarized radiation of sources through
the ICM. Recently, Newman, Newman, \& Rephaeli (2002) have pointed
out that many and large uncertainties are associated to the
determination of $B_{FR}$ (see also Govoni \etal 2002). However,
this discrepancy can be attenuated considering models that include
the effects of more realistic electron spectra, spatial profiles
of the magnetic fields and anisotropies in the pitch angle
distribution of the electrons (Goldshmidt \& Rephaeli 1993;
Brunetti \etal 2001; Petrosian 2001). Another possibility is that
the detected excesses are due to the presence of very obscured
sources like Circinus (Matt \etal 1999) in the field of view of
the PDS (1.3$^{\circ}$). In the case of Coma and A2256 the
analysis of the MECS images excludes the presence of this kind of
sources in the central region ($\sim$ 30$'$ in radius) of the
cluster unless the obscured source is within 2$'$ of the central
bright cores. We have re-observed A2256 after about two years from
the first observation and the two spectra are consistent
(Fusco-Femiano \etal 2002). Besides, both observations comprise
two exposures with a time interval of $\sim$ 1 year and $\sim$ 1
month, respectively and all these observations do not show
significant flux variations. These results and the fact that the
two clusters, Coma and A2256, with a detected HXR excess both have
extended radio regions strongly supports the idea that a diffuse
nonthermal mechanism involving the ICM is responsible for the
observed excesses. Support to this interpretation is given by a
second recent \rxte observation of the Coma cluster (Rephaeli \&
Gruber 2002) that confirms the previous observation (Rephaeli,
Gruber, \& Blanco 1999) of a likely presence of a nonthermal
component in the spectrum of the cluster.

For A754 we discuss two possible origins for the detected excess
at the level of $\sim 3.2\sigma$, lower than that measured in Coma
and A2256 ($\sim 4.5\sigma$). As reported in the Introduction, the
ICM of A754 appears to be a suitable place to host relativistic
electrons considering that X-ray and optical observations indicate
that the cluster is undergoing a violent merger process, and that
Hydro/N-body simulations show that the merger must be very recent.
The time elapsed from the start of the merger in A754 ($<$ 0.3
Gyr) is not greater than the radiative lifetime for electrons with
$\gamma \sim 10^4$. Shocks and turbulence associated with a recent
or current major merger event can be able to provide energy for
particle acceleration and magnetic field amplification to produce
diffuse radio emission (Tribble 1993; Brunetti \etal 2001). This
scenario seems to be confirmed by the presence of diffuse radio
emission (Kassim \etal 2001) and now definitely confirmed by our
deeper VLA observation at higher resolution at 1.4 GHz. The
nonthermal HXR detected by the PDS could be IC emission by the
same radio electrons scattering CMB photons. The lack of
variability in the nonthermal flux detected by this long \bsax
observation is compatible with a diffuse mechanism of the observed
excess. Assuming that the measured \nt flux of $\sim 10^{-11}\erg$
in the 40-100 keV energy range is due to IC emission, the
volume-averaged intracluster magnetic field is $\sim 0.1 \mu G$
obtained combining the radio halo and X-ray fluxes (see
Fusco-Femiano \etal 1999, eq.1) and using our measured value of
1.5 for the radio halo index. A similar value of $B_X$ is obtained
for the radio relic assuming a radio spectral index of 1.7. The
equipartition value of the magnetic field, computed with standard
assumptions, is in the range $0.3-0.6 \mu G$.

The presence of diffuse radio radiation in A754, considered the
prototype of a merging cluster, is a further evidence of the link
between Mpc-scale radio emission and very recent or current merger
processes (Feretti 1999). Radio observations and the detection of
nonthermal HXR emission are of great importance for the
understanding of the origin of the electron population responsible
for nonthermal phenomena in clusters of galaxies. In the Coma
cluster, the prototype of halo clusters, the spectral cut-off
(Deiss \etal 1997) represents a strong indication for the presence
of a cut-off in the emitting electron spectrum. This cut-off and
the HXR excess detected in the Coma cluster may be naturally
accounted for in the context of re-acceleration models (Brunetti
\etal 2001) in which relativistic electrons with $\gamma\sim$
100-300, produced by several sources in galaxy clusters and
accumulated in the cluster volume over cosmological time, can be
re-accelerated by various processes at the energies required to
explain the radio halo in the Coma cluster. This cut-off is not
naturally expected if the radio emission is produced by secondary
electrons (Brunetti 2002) due to the decay of charged pions
generated in cosmic ray collisions (Dennison 1980; Blasi \&
Colafrancesco 1999; Dolag \& Ensslin 2000; Miniati \etal 2001).
The radio and HXR observations of A754 seem to confirm the
scenario described by the re-acceleration models in the Coma
cluster. The detection of the hard excess in A754 determines, in
the framework of the IC model, a volume-averaged intracluster
magnetic field of the same order ($B_X\sim 0.1\mu G$) of that
determined in the Coma cluster (Fusco-Femiano \etal 1999). This
value of $B_X$ implies relativistic electrons at energies
$\gamma\sim 10^4$ to explain the observed diffuse synchrotron
emission. At these energies IC losses may determine a cut-off in
the electron spectrum in agreement with the spectral cut-off
observed in A754. So, the radio and HXR results obtained by our
observations of A754 support the scenario that primary and not
secondary electrons are responsible for nonthermal emission in
clusters of galaxies.

However, we cannot ignore the presence of the radio galaxy 26W20
in the field of view of the PDS, located at a distance of about
27$'$ from the \bsax pointing, that could be responsible for the
hard excess in A754. This source, discovered in the Westerbork
radio survey by Harris \etal (1980), shows a X-ray bright core
similar to that of a BL Lac object but with weak emission lines
and this makes 26W20 different from a typical BL Lac object. The
source is preferred to be classified as tailed radio galaxy for
the presence of a single jet or tail and being member of a small
group of galaxies. The radio galaxy has had several X-ray
observations due to its close proximity to A754. It was first
observed with \einstein (Harris, Costain \& Dewdney 1984) and the
IPC spectrum could be equally well fit by a power law (energy
index 0.8$\pm$0.4) or with a thermal model (kT $\sim$ 3 keV). In
the following the nonthermal nature of the X-ray emission was
clarified by a \rosat observation. In particular, the fit to an
on-axis PSPC observation gives a core emission described by a
nonthermal model with energy index 1.32$\pm$0.17 and log $N_H =
20.90\pm 0.04$. The X-ray luminosity is $\sim 3\times
10^{43}\ergs$ in the 0.5 - 3 KeV energy range (Silverman, Harris,
\& Junor 1998). The source shows variability. An 18\% increase in
the luminosity has been determined by two PSPC observations within
5 days in 1992.

The spectral energy distributions (SED) of BL Lac sources are
characterized by two main peaks. The peak at low energy is
commonly explained as synchrotron emission by relativistic
electrons while that at higher energies as IC emission of the same
electrons scattering the synchrotron photons (e.g. Dermer \etal
1997; Ghisellini \etal 1998).

\begin{figure*}
\centering
  \resizebox{10cm}{!}{\includegraphics[angle=-90]{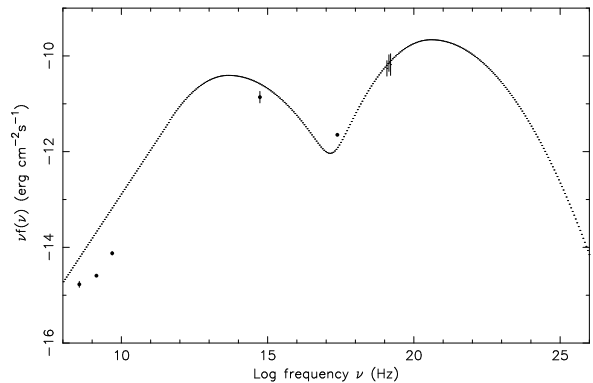}}
  \hfill
  \parbox[b]{80mm}{\caption{Spectral energy distribution for 26W20. The first and third radio
  points
   are taken from NED, the second from NVSS. The optical point from the USNO catalog and the ROSAT point from
   the PSPC observation by Silverman, Harris, \& Junor (1998). The highest energy points
  refer to the hard excess detected by the PDS observation. The dotted line is the fit to the SED with a SSC model.}
   \label{pippo2}}
\end{figure*}

In Figure 3 we show the SED for 26W20 where the highest energy
points refer to the PDS observation assuming that this source is
responsible for the detected excess. The fit with a Synchrotron
Self-Compton (SSC) model is consistent with the SED of a BL Lac
object (Perry 2001; Padovani, Perlman, \& Landt 2002; Giommi \etal
2002), but a flat energy index of about 0.3 is required to
extrapolate the flux detected by \rosat in the PDS energy range,
taking into account the angular response of the detector. The
inclusion of the PDS points makes difficult to fit well all the
points of the SED. However, due to the variability of this kind of
sources only simultaneous observations can be used to fix the
spectral properties in a large energy band. Unfortunately, the
source does not appear in the field of view of the MECS because
hidden by one of the calibration sources of the instrument.

\section{Conclusions}

Our VLA observation at 1.4 GHz definitely confirms diffuse radio
emission in A754 composed by a radio halo and a relic, although a
connection relic-halo cannot be excluded. The relic image at 1.4
GHz appears much more extended than that detected at 74 MHz by
Kassim \etal (2001). Our observation determines with accuracy the
slope of the radio halo spectrum in the 0.3-1.4 GHz frequency
range that results to be steeper than that obtained in the band
74-330 MHz, indicating the presence of a radio spectral cutoff. A
\bsax observation of A754 shows a nonthermal HXR excess at
energies above $\sim$ 45 keV with respect to the thermal emission,
at a confidence level of $\sim 3\sigma$. One possible explanation
is IC emission for the same relativistic electrons responsible for
the diffuse radio emission. This interpretation has been invoked
to explain the \nt emission detected in Coma and A2256 both
showing extended radio regions. The lack of significative
variability of the \nt flux detected in A754 is consistent with a
diffuse mechanism for the observed excess. This detection, that
needs to be confirmed by future observations, and the radio
results seem indicate that primary re-accelerated electrons and
not secondary electrons are at the origin of the observed
nonthermal emission in clusters of galaxies. At present, it exists
also the possibility that the radio galaxy 26W20, located in the
field of view of the PDS, is responsible for the reported hard
excess. The extrapolation of the \rosat flux in the PDS energy
range requires a flat index to explain the nonthermal HXR flux.
However, considering the non simultaneity of the \rosat and \bsax
observations the SED may have assumed a different shape at the
moment of the \bsax observation. To discriminate between these two
interpretations a hard X-ray observation with imaging instruments
is necessary.

\begin{acknowledgements}
We thank G.Brunetti and P.Giommi for useful suggestions regarding
the interpretation of the results. We are grateful to the referee
for the stimulating comments.
\end{acknowledgements}


\begin{thebibliography}{}
\bibitem[]{}
Bird, C.M. 1994, AJ, 107, 1637
\bibitem[]{}
Blasi, P. 2000, ApJ, 525, 603
\bibitem[]{}
Blasi, P. \& Colafrancesco, S. 1999, Astropart. Phys., 12, 169
\bibitem[]{}
Bliton, M., Rizza, E., Burns, J.O., Owen, F.N., Ledlow, M.J. 1998,
MNRAS, 301, 609
\bibitem[]{}
Brunetti, G., Setti, G., Feretti, L., \& Giovannini, G. 2001,
MNRAS, 320, 365
\bibitem[]{}
Brunetti, G. 2002, invited review in the Meeting: "Matter and
Energy in Clusters of Galaxies", ASP Conf.Ser., eds: S.Bowyer \&
C-Y. Hwang; astro-ph/0208074
\bibitem[]{}
Chiappetti, L. \& Dal Fiume, D. 1997, in Data Analysis in
Astronomy, ed. V. Di Gesu` \etal (Singapore: World Scientific),
101
\bibitem[]{}
Dal Fiume, D. Frontera, F., Nicastro, L., Orlandini, M., Palazzi,
E., Costa, E., Feroci, M., \& Zavattini, G. 1997, in Data Analysis
in Astronomy, ed. V. Di Gesu` \etal (Singapore: World Scientific),
111
\bibitem[]{}
De Grandi, S. \& Molendi, S. 2001, ApJ, 551, 153
\bibitem[]{}
Dennison, B. 1980, ApJ, 239, 93
\bibitem[]{}
Dermer, C.D., Sturner, S.J., \& Schlickeiser, R. 1997, ApJS, 109,
103
\bibitem[]{}
Deiss, B.M., Reich, W., Lesch, H.,\& Wielebinski, R. 1997, A\&A,
321, 55
\bibitem[]{}
Dennison, B. 1980, ApJ, 239, L93
\bibitem[]{}
Dogiel, V.A. 2000, A\&A, 357, 66
\bibitem[]{}
Dolag, K., \& Ensslin, T.A. 2000, A\&A, 362, 151
\bibitem[]{}
Dressler, A. 1980, ApJS, 42, 565
\bibitem[]{}
Feretti, L., Dallacasa, D., Giovannini, G., \& Tagliani, A. 1995,
A\&A, 302, 680
\bibitem[]{}
Feretti, L. 1999, in Diffuse Thermal and Relativistic Plasma in
Galaxy Clusters, ed. H. Bohringer, L. Feretti, \& P. Schuecker
(MPE Rep. 271; Garching: MPE), 3
\bibitem[]{}
Frontera, F., Costa, E., Dal Fiume, D., Feroci, M., Nicastro, L.,
Orlandini, M., Palazzi, E., \& Zavattini, G. 1997, A\&AS, 122, 357
\bibitem[]{}
Fusco-Femiano, R., Dal Fiume, D., Feretti, L., Giovannini, G.,
Grandi, P., Matt, G., Molendi, S., \& Santangelo, A. 1999, ApJ,
513, L21
\bibitem[]{}
Fusco-Femiano, R., Dal Fiume, D., De Grandi, S., Feretti, L.,
Giovannini, G., Grandi, P., Malizia, A., Matt, G., \& Molendi, S.
2000, ApJ, 534, L10
\bibitem[]{}
Fusco-Femiano, R., Dal Fiume, D., Orlandini, M., Brunetti, G.,
Feretti, L., \& Giovannini, G. 2001, ApJ, 552, L100
\bibitem[]{}
Fusco-Femiano, R. \etal 2002, to appear in the Proc. "Matter and
Energy in Clusters of Galaxies", April 2002, Taiwan, ASP, Conf.
Ser., eds.: S.Bowyer \& C-Y.Hwang; astro-ph/0207241
\bibitem[]{}
Ghisellini, G., Celotti, A., Fossati, G., Maraschi, L., \&
Comastri, A. 1998, MNRAS, 301, 451
\bibitem[]{}
Giommi, P., Massaro, E., \& Palumbo, G. 2002, Proc. of Workshop
"Blazar Astrophysics with \bsax and Other Observatories", December
10-11, 2001, Frascati, Italy, in press
\bibitem[]{}
Giovannini, G., Feretti, L., Venturi, T., Kim, K.T., \& Kronberg,
P.P. 1993, ApJ, 406, 399
\bibitem[]{}
Giovannini, G., Tordi, M., \& Feretti, L. 1999, New Astronomy, 4,
141
\bibitem[]{}
Govoni, F., Feretti, L., Giovannini, G., Murgia, M., Taylor, G.,
\& Dallacasa, D. 2002, to appear in the Proc. of "Matter and
Energy in Clusters of Galaxies", 23-27 April, Chung-Li, Taiwan,
eds. S.Bowyer \& C.-Y.Hwang.
\bibitem[]{}
Goldshmith, O. \& Rephaeli, Y. 1993, ApJ, 411, 518
\bibitem[]{}
Harris, D.E. \etal 1980, A\&A, 90, 283
\bibitem[]{}
Harris, D.E., Costain, C.H., \& Dewdney, P.E. 1984, ApJ, 280, 532
\bibitem[]{}
Henriksen, M.J. \& Markevitch, M.L. 1996, ApJ, 466, L79
\bibitem[]{}
Henry, J.P. \& Briel, U.G. 1995, ApJ, 443, L9
\bibitem[]{}
Kassim, N.E., Clarke, T.E., En$\ss$lin, T.A., Cohen, A.S., \&
Neumann, D.M. 2001, ApJ, 559, 785
\bibitem[]{}Liang, H., Dogiel, V.A., \& Birkinshaw, M. 2002,
MNRAS, in press; astro-ph/0208509
\bibitem[]{}
Markevitch, M., Forman, W.R., Sarazin, C.L.
\bibitem[]{}
Matt, G. \etal 1999, A\&A, 341, L39
\bibitem[]{}
Miniati, F., Jones, T.W., Kang, H., \& Ryu, D. 2001, ApJ, 562, 233
\bibitem[]{}
Newman, W.I., Newman, A.L., \& Rephaeli, Y. 2002, astro-ph/0204451
\bibitem[]{}
Padovani, P., Perlman, E.S., \& Landt, H. 2002, in preparation
\bibitem[]{}
Perry, M. 2001, PhD thesis
\bibitem[]{}
Petrosian, V. 2001, ApJ, 557, 560
\bibitem[]{}
Rephaeli, Y., Gruber, D.E., \& Blanco, P. 1999, ApJ, 511, L21
\bibitem[]{}
Rephaeli, Y., \& Gruber, D. 2002, astro-ph/0207443
\bibitem[]{}
Roettiger, K., Stone, J.M., \& Mushotzky, R.F. 1998, ApJ, 493, 62
\bibitem[]{}
Roettiger, K., Stone, J.M., \& Burns, J.O. 1999, ApJ, 518, 594
\bibitem[]{}
Roettiger, K., Burns, J.O., \& Stone, J.M. 1999, ApJ, 518, 603
\bibitem[]{}
Sarazin, C.L. \& Kempner, J.C. 2000, ApJ, 533, 73
\bibitem[]{}
Sijbring, L.G. 1993, PhD Thesis, Groeningen Univ.
\bibitem[]{}
Silverman, J.D., Harris, D.E., \& Junor, W. 1998, A\&A, 335, 443
\bibitem[]{}
Tribble, P.C. 1993, MNRAS, 261, 57
\bibitem[]{}
Valinia, A., Henriksen, M.J., Loewenstein, M., Roettiger, K.,
Mushotzky, R.F., \& Madejski, G. 1999, ApJ, 515, 42
\bibitem[]{}
Zabludoff, A.I. \& Zaritsky, D. 1995, ApJ, 447, L21

\end{thebibliography}
\end{document}